# A family of fast-decodable MIDO codes from crossed-product algebras over $\mathbb{Q}$

Laura Luzzi
Alcatel-Lucent Chair on Flexible Radio
Supélec
91192 Gif-sur-Yvette, France
Email: laura.luzzi@supelec.fr

Frédérique Oggier
Division of Mathematical Sciences
School of Physical and Mathematical Sciences
Nanyang Technological University, Singapore
Email: frederique@ntu.edu.sg

*Abstract*—Multiple Input Double Output (MIDO) asymmetric space-time codes for $4$ transmit antennas and $2$ receive antennas can be employed in the downlink from base stations to portable devices. Previous MIDO code constructions with low Maximum Likelihood (ML) decoding complexity, full diversity and the non-vanishing determinant (NVD) property are mostly based on cyclic division algebras.
In this paper, a new family of MIDO codes with the NVD property based on *crossed-product algebras* over $\mathbb{Q}$ is introduced. Fast decodability follows naturally from the structure of the codewords which consist of four generalized Alamouti blocks. The associated ML complexity order is the lowest known for full-rate MIDO codes ($O(M^{10})$ instead of $O(M^{16})$ with respect to the real constellation size $M$). Numerical simulations show that these codes have a performance from comparable up to 1dB gain compared to the best known MIDO code with the same complexity.

## I. INTRODUCTION

There are many wireless channel scenarios where the number of antennas is asymmetric, in particular, where the transmitter has many antennas, while the receiver, for example being a portable device, has few of them. MIDO channels, which stands for Multiple Inputs Double Outputs, refer to such systems. The model we will consider in this paper is a MIDO coherent Rayleigh fading channel, with $4$ antennas at the transmitter and $2$ antennas at the receiver, which furthermore has perfect channel state information at the receiver.

Since the computing power available at the receiver is typically very limited, the design of MIDO codes must take into account the decoding complexity order. A code is called *fast-decodable* if the research tree in the sphere decoding algorithm [10] can be simplified. In general, the complexity order of the real sphere decoding algorithm for a system with $m$ transmit antennas and $n$ receive antennas employing a space-time code and real signal constellations of size $M$ is $O(M^{2mn})$.

### A. Related work

The first low ML complexity MIDO code was proposed in [2]. This code has complexity order $O(M^{12})$ instead of $O(M^{16})$; it is not full-rank but still achieves good performance for moderate values of SNR. A full-rank MIDO code with $O(M^{10})$ complexity and high coding gain which is conjectured to have the non-vanishing determinant (NVD) property was presented in [9]. Recently, MIDO codes with the NVD property based on cyclic division algebras with complexity order up to $O(M^{10})$ were also constructed in [7, 6, 11].

In this paper, we consider an alternative approach and introduce a new family of $O(M^{10})$ decoding complexity MIDO codes with the NVD property based on *crossed-product algebras* over $\mathbb{Q}$. Crossed-product algebras were already used to construct one example of fast-decodable MIDO code in [7]; however, this example is based on puncturing a $4 \times 4$ full-rate space-time code and its performance is not very good. The constructions presented here are tailored for the $4 \times 2$ case and do not require any puncturing.

### B. Alamouti-like structures

Let $(\ )^*$ denote the complex conjugation for a scalar, and the Hermitian transpose for a vector or a matrix. Recall that an Alamouti block code is given by

$$\begin{pmatrix} y_1 & -y_2^* \\ y_2 & y_1^* \end{pmatrix},$$

with $y_1, y_2 \in \mathbb{C}$. It has the property that its columns are orthonormal. This property allows fast decoding, and consequently many of the attempts to construct fast decodable space-time codes have tried to mimic it. In this paper we will consider generalized Alamouti codewords of the form

$$\begin{pmatrix} y_1 & -\alpha y_2^* \\ \alpha^* y_2 & y_1^* \end{pmatrix}, \ \alpha \in \mathbb{C} \qquad (1)$$

which will make the columns orthogonal (furthermore orthonormal if $|\alpha| = 1$). Note that this definition is different from the one in [2]. Note also that if we have a matrix of the form

$$\begin{pmatrix} y_1 & -\alpha y_2^* \\ y_2 & y_1^* \end{pmatrix}, \ \alpha \in \mathbb{R}, \ \alpha > 0, \qquad (2)$$

then by multiplying the second column by $1/\sqrt{\alpha}$, and the second row by $\sqrt{\alpha}$, we get, without changing the matrix determinant,

$$\begin{pmatrix} y_1 & -\sqrt{\alpha} y_2^* \\ \sqrt{\alpha} y_2 & y_1^* \end{pmatrix}, \qquad (3)$$

a particular case of (1) when $\alpha$ is real. As far as decoding is concerned, it is enough to ask for the columns to be orthogonal; the orthonormality does not improve the decoding complexity, but rather the performance by ensuring that the energy is balanced across time and antennas.

Our goal in this paper is to construct a code carrying 8 complex information symbols of the form

$$\begin{pmatrix} A & C \\ B & D \end{pmatrix}, \quad (4)$$

where the $2 \times 2$ blocks $A$ and $D$ are generalized Alamouti codes of the form (1) (it will be discussed in Section IV why the focus is on the blocks $A$ and $D$), preferably with columns as close to orthonormal as possible. This means that the energy of the symbols might not be balanced, as in (3).

## II. THE FRAMEWORK OF CROSSED PRODUCT ALGEBRAS

The incentive to consider biquadratic crossed product algebras as underlying algebraic structure to construct fast decodable MIDO codes is two-fold: first, we will see below that the representation of these algebras naturally gives rise to codewords of the form (4), and furthermore, as is the case with traditional space-time coding using division algebras, a codebook with full diversity is obtained from division crossed product algebras.

### A. Crossed product algebras of degree 4

We will consider as in [1] a crossed product algebra $\mathcal{A} = (L/K, a, b, u)$ over the biquadratic extension $L/K$, where $L = K(\sqrt{d}, \sqrt{d'})$, and

$$\langle \sigma \rangle = Gal(K(\sqrt{d'})/K), \ \langle \tau \rangle = Gal(K(\sqrt{d})/K).$$

Such an algebra is of the form $\mathcal{A} = L \oplus eL \oplus fL \oplus efL$, where $e^2 = a \in K(\sqrt{d})$, $f^2 = b \in K(\sqrt{d'})$, $xe = e\sigma(x) \ \forall x \in L$, $xf = f\tau(x) \ \forall x \in L$, and $fe = efu$ for $u$ a non-zero element of $L$ such that $u\sigma(u) = a/\tau(a)$, $u\tau(u) = \sigma(b)/b$. Elements of $\mathcal{A}$ admit the following matrix representation:

$$\begin{pmatrix} x_0 & a\sigma(x_1) & b\tau(x_2) & ab\tau(u)\sigma\tau(x_3) \\ x_1 & \sigma(x_0) & b\tau(x_3) & b\tau(u)\sigma\tau(x_2) \\ x_2 & \tau(a)u\sigma(x_3) & \tau(x_0) & \tau(a)\sigma\tau(x_1) \\ x_3 & u\sigma(x_2) & \tau(x_1) & \sigma\tau(x_0) \end{pmatrix}, \quad (5)$$

with $x_0, x_1, x_2, x_3 \in L$.
As shown in [1], $u$ is such that $N_{L/K}(u) = 1$, and suitable $a$ and $b$ are determined by the choice of $u$ in the following way:

$$a = \begin{cases} k\sqrt{d} & \text{if } u\sigma(u) = -1, \ k \in K; \\ l(1 + u\sigma(u)) & \text{otherwise, } l \in K. \end{cases} \quad (6)$$

In the latter case, in order to have a non-degenerate algebra, $u\sigma(u)$ should belong to $K(\sqrt{d}) \smallsetminus K$. Similarly,

$$b = \begin{cases} k'\sqrt{d'} & \text{if } u\tau(u) = -1; \ k' \in K \\ \frac{l'}{1+u\tau(u)} & \text{otherwise, } l' \in K, \end{cases} \quad (7)$$

with $u\tau(u) \in K(\sqrt{d'}) \smallsetminus K$. However, checking whether the resulting crossed product algebra $\mathcal{A}$ is a division algebra only depends on the choice of $u$:

**Theorem 1.** *[1] Let $K$ be a number field, and let $\mathcal{A} = (L/K, a, b, u)$ be a crossed product algebra. Then $\mathcal{A}$ is a division algebra if and only if*

1) $\left( \frac{-d, d'}{K} \right)$ *is a division algebra and* $u\sigma(u) = -1$,
2) $\left( \frac{d', 2+Tr_{K(\sqrt{d})/K}(u\sigma(u))}{K} \right)$ *is a division algebra and* $u\sigma(u) \neq -1$.

There is a similar equivalent formulation depending on whether $u\tau(u) = -1$. Since we need 8 complex symbols, it is enough to consider a crossed product algebra $\mathcal{A}$ over a biquadratic extension $L$ of $K = \mathbb{Q}$. Such an algebra is of index 4, thus we can encode 16 real symbols, that is 8 complex ones. In order to obtain a matrix of the form (4) from (5),

$$A = \begin{pmatrix} x_0 & a\sigma(x_1) \\ x_1 & \sigma(x_0) \end{pmatrix}, \quad D = \begin{pmatrix} \tau(x_0) & \tau(a)\sigma\tau(x_1) \\ \tau(x_1) & \sigma\tau(x_0) \end{pmatrix}$$

should be generalized Alamouti blocks. Alternatively, by swapping the second and third rows and the second and third columns in (5), we get

$$A = \begin{pmatrix} x_0 & b\tau(x_2) \\ x_2 & \tau(x_0) \end{pmatrix}, \quad D = \begin{pmatrix} \sigma(x_0) & b\tau(u)\sigma\tau(x_2) \\ u\sigma(x_2) & \sigma\tau(x_0) \end{pmatrix}.$$

Finally, by swapping the second and fourth rows and columns, we obtain

$$A = \begin{pmatrix} x_0 & ab\tau(u)\sigma\tau(x_3) \\ x_3 & \sigma\tau(x_0) \end{pmatrix}, \quad D = \begin{pmatrix} \tau(x_0) & \tau(a)u\sigma(x_3) \\ b\tau(x_3) & \sigma(x_0) \end{pmatrix} \quad (8)$$

There are three possible choices for $K(\sqrt{d}) = \mathbb{Q}(\sqrt{d})$ and $K(\sqrt{d'}) = \mathbb{Q}(\sqrt{d'})$:

- both are imaginary quadratic fields,
- both are real quadratic fields,
- one is a real quadratic field, the other is an imaginary field.

Since typical signal constellations such as QAM are encoded using $\mathbb{Q}(i)$, we will assume that one of the two quadratic fields is $\mathbb{Q}(i)$ and will thus not consider the case of two real quadratic fields. Due to the lack of space, we will focus on the case of two imaginary quadratic fields, where $\sigma\tau$ acts as the complex conjugation, and choose codewords of the form (8).

### B. Crossed product algebras over imaginary fields

Consider now the case where $d = -c < 0, d' = -c' < 0$, that is $L = \mathbb{Q}(\sqrt{-c}, \sqrt{-c'})$. We are mostly interested in the case when $c' = 1$, but nevertheless will show later that the construction is also possible for other values of $c'$.

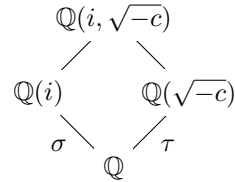

The Galois group of $\mathbb{Q}(i)/\mathbb{Q}$, resp. $\mathbb{Q}(\sqrt{-c})/\mathbb{Q}$ is denoted by $\langle \sigma \rangle$, resp. $\langle \tau \rangle$ with $\sigma(i) = -i$, $\tau(\sqrt{-c}) = -\sqrt{-c}$. Every element of $L$ is of the form $x = a_1 + a_2 i + a_3 \sqrt{-c} + a_4 i \sqrt{-c}$,

$a_1, a_2, a_3, a_4 \in \mathbb{Q}$, and we extend $\sigma$ and $\tau$ to $L$ so as to get

$$\sigma(x) = a_1 - a_2 i + a_3 \sqrt{-c} - a_4 i \sqrt{-c},$$
$$\tau(x) = a_1 + a_2 i - a_3 \sqrt{-c} - a_4 i \sqrt{-c},$$
$$\sigma\tau(x) = a_1 - a_2 i - a_3 \sqrt{-c} + a_4 i \sqrt{-c} = x^*.$$

We need to obtain two properties:
- the Alamouti-like block structure for fast-decodability,
- the algebra should preferably be a division algebra to guarantee a good behaviour at high SNR.

Let us first exploit the property $\sigma\tau(x) = x^*$ to construct a code with an Alamouti block structure by swapping the second and fourth row and the second and fourth column of the representation (5), as already mentioned in (8):

$$\begin{pmatrix} x_0 & ab\tau(u)\sigma\tau(x_3) & b\tau(x_2) & a\sigma(x_1) \\ x_3 & \sigma\tau(x_0) & \tau(x_1) & u\sigma(x_2) \\ x_2 & \tau(a)\sigma\tau(x_1) & \tau(x_0) & \tau(a)u\sigma(x_3) \\ x_1 & b\tau(u)\sigma\tau(x_2) & b\tau(x_3) & \sigma(x_0) \end{pmatrix}.$$

Since complex conjugation commutes with $\sigma$ and $\tau$, we can rewrite it as

$$\begin{pmatrix} x_0 & ab\tau(u)x_3^* & b\tau(x_2) & a\tau(x_1)^* \\ x_3 & x_0^* & \tau(x_1) & u\tau(x_2)^* \\ x_2 & \tau(a)x_1^* & \tau(x_0) & \tau(a)u\tau(x_3)^* \\ x_1 & b\tau(u)x_2^* & b\tau(x_3) & \tau(x_0)^* \end{pmatrix}. \quad (9)$$

We are now left with the choice of $a, b$ and $u$. From (2), we have as condition on these parameters that

$$ab\tau(u) \in \mathbb{R}, \ ab\tau(u) < 0. \quad (10)$$

To obtain a suitable crossed product algebra $\mathcal{A}$, we need to choose the element $u$ (with $N_{L/K}(u) = 1$), and take $a,b$ such that

$$a = \begin{cases} k\sqrt{-c}, \ k \in \mathbb{Q} & \text{if } u\sigma(u) = -1; \\ l(1 + u\sigma(u)), \ l \in \mathbb{Q} & \text{otherwise, } u\sigma(u) \notin K, \end{cases} \quad (11)$$

and

$$b = \begin{cases} k'i, \ k' \in \mathbb{Q} & \text{if } u\tau(u) = -1; \\ \frac{l'}{1+u\tau(u)}, \ l' \in \mathbb{Q} & \text{otherwise, } u\tau(u) \notin K. \end{cases} \quad (12)$$

For such an algebra to be a division algebra, when $u\sigma(u) = -1$, we have by Theorem 1 that it is enough to check whether $\left(\frac{c,-1}{\mathbb{Q}}\right)$ is. This in turn is equivalent to see whether $c = N_{\mathbb{Q}(i)/\mathbb{Q}}(s)$ for some $s \in \mathbb{Q}(i)$. Since $N_{\mathbb{Q}(i)/\mathbb{Q}}(s) = s_1^2 + s_2^2$ for $s = s_1 + is_2$, $s_1, s_2 \in \mathbb{Q}$, we finally need to check whether $c$ can be written as a sum of two squares. We also consider the case where $c' = 2$, which is the next smallest $c'$ after $c' = 1$. Similarly in this case when $u\sigma(u) = -1$, we need to check whether $\left(\frac{c,-2}{\mathbb{Q}}\right)$ is a division algebra, that is, whether $c = s_1^2 + 2s_2^2$, $s_1, s_2 \in \mathbb{Q}$. Recall that $c$ can be written as the sum of two rational squares if and only if all its odd prime factors which are congruent to 3 (mod 4) occur to an even exponent; similarly, $c$ can be written in the form $s_1^2 + 2s_2^2$ if and only if its odd prime factors which are congruent to 5 or 7 (mod 8) occur to an even exponent. The first smallest possible values for $c$ are listed below, for both $c' = 1$ and $c' = 2$ [5].

| $(c, -c')$ | division algebra | $(c, -c')$ | division algebra |
|---|---|---|---|
| (2,-1) | no (2 = 1 + 1) | (2,-2) | no (2 = 0 + 2) |
| (3,-1) | yes | (3,-2) | no (3 = 1 + 2) |
| (5,-1) | no (5 = 1 + 4) | (5,-2) | yes |
| (6,-1) | yes | (6,-2) | no (6 = 4 + 2) |
| (7,-1) | yes | (7,-2) | yes |
| (10,-1) | no (10 = 9 + 1) | (10,-2) | yes |
| (11,-1) | yes | (11,-2) | no (11 = 9 + 2) |
| (13,-1) | no (13 = 9 + 4) | (13,-2) | yes |

The condition when $u\sigma(u) \neq -1$ is a priori less systematic to check, since it relies on the trace of $u\sigma(u)$, though we can look for an element $u$ such that the trace of $u\sigma(u)$ is $c - 2$, where $c$ is such that $(c, -c')$ is a division algebra.

We finally discuss briefly how to find $u$ with $N_{L/K}(u) = 1$. A natural choice is to start by taking $u$ a unit in $L$. Recall that by Dirichlet's unit theorem, the units of an algebraic number field $L$ are a multiplicative group generated by a set of *fundamental units*. The number of fundamental units is $r_1 + r_2 - 1$, where $r_1$ is the number of real embeddings of $L$, and $r_2$ is the number of pairs of complex embeddings.

In the case $L = \mathbb{Q}(\sqrt{-c}, \sqrt{-c'})$, we have $r_1 = 0$ and $r_2 = 2$, therefore there is only one fundamental unit.

## III. CODE CONSTRUCTIONS

### A. A generic construction

Suppose that there exists $u$ in $L$ with $N_{L/K}(u) = 1$, and that the corresponding $a$ and $b$ as defined by (11) and (12) satisfy the following conditions:

$$u\sigma(u) = -1, \ u\tau(u) = \varepsilon \in \{-1, i, -i\},$$
$$ab\tau(u) \in \mathbb{R}, \ ab\tau(u) < 0. \quad (13)$$

We now set

$$\alpha = -ab\tau(u) > 0.$$

Observe that the first condition implies that $a = i\sqrt{c}k$, and take $k = 1$. Since $b \in K(\sqrt{-c'}) = \mathbb{Q}(i)$, $\sigma(b) = b^*$ and the second condition, with $l' = 1$, implies that $\varepsilon = u\tau(u) = \frac{b^*}{b}$. We then have

$$b\tau(u) = -\frac{\alpha}{a} = \frac{i\alpha}{\sqrt{c}},$$
$$u = \frac{\varepsilon}{\tau(u)} = -\frac{\varepsilon ab}{\alpha} = -\frac{\varepsilon bi\sqrt{c}}{\alpha} = -\frac{i\sqrt{c}b^*}{\alpha},$$
$$\tau(a)u = -i\sqrt{c}u = -\frac{cb\varepsilon}{\alpha} = -\frac{cb^*}{\alpha},$$

where the second equality uses the expression computed for $\tau(u)$ above. Replacing in the expression (9), we find that a codeword is of the form

$$\begin{pmatrix} x_0 & -\alpha x_3^* & b\tau(x_2) & i\sqrt{c}\tau(x_1)^* \\ x_3 & x_0^* & \tau(x_1) & -\frac{i\sqrt{c}b^*}{\alpha}\tau(x_2)^* \\ x_2 & -i\sqrt{c}x_1^* & \tau(x_0) & -\frac{cb^*}{\alpha}\tau(x_3)^* \\ x_1 & \frac{i\alpha}{\sqrt{c}}x_2^* & b\tau(x_3) & \tau(x_0)^* \end{pmatrix}.$$

Dividing the first row by $\sqrt{\alpha}$ and multiplying the first column by $\sqrt{\alpha}$, and further multiplying the fourth column by $\frac{\sqrt{\alpha}}{\sqrt{c}}$ and dividing the fourth row by $\frac{\sqrt{\alpha}}{\sqrt{c}}$, yields:

$$\begin{pmatrix} x_0 & -\sqrt{\alpha}x_3^* & \frac{b}{\sqrt{\alpha}}\tau(x_2) & i\tau(x_1)^* \\ \sqrt{\alpha}x_3 & x_0^* & \tau(x_1) & -\frac{ib^*}{\sqrt{\alpha}}\tau(x_2)^* \\ \sqrt{\alpha}x_2 & -i\sqrt{c}x_1^* & \tau(x_0) & -\frac{b^*\sqrt{c}}{\sqrt{\alpha}}\tau(x_3)^* \\ \sqrt{c}x_1 & i\sqrt{\alpha}x_2^* & \frac{b\sqrt{c}}{\sqrt{\alpha}}\tau(x_3) & \tau(x_0)^* \end{pmatrix}. \quad (14)$$

We have thus obtained a codeword composed of four generalized Alamouti blocks:

$$\begin{pmatrix} z_1 & -z_2^* & z_5 & iz_6^* \\ z_2 & z_1^* & z_6 & -iz_5^* \\ z_3 & -iz_4^* & z_7 & -z_8^* \\ z_4 & iz_3^* & z_8 & z_7^* \end{pmatrix}.$$

We now provide examples of such code constructions, with values of $c$ which give a division algebra, namely $c' = 1$ and $c = 3, 6$ and $11$. In the following, the fundamental units have been computed using the KASH software [4].

**Example 1** ($L = \mathbb{Q}(i, \sqrt{3})$). Let $c' = 1$, $c = 3$. Then $L = \mathbb{Q}(i, \sqrt{3}) = \mathbb{Q}(\zeta_{12})$. The fundamental unit of $L$ is $v = \left(\frac{1+i}{2}\right)(\sqrt{3}-1)$. We will choose $u = \tau(v) = \left(\frac{1+i}{2}\right)(-\sqrt{3}-1)$. We have $u\sigma(u) = -1$, $u\tau(u) = -i$. Then $a = \sqrt{-3}$ and $b = \frac{1+i}{2}$. We have $ab\tau(u) = \sqrt{3}\left(\frac{1-\sqrt{3}}{2}\right) < 0$, so the conditions (13) are satisfied.

**Example 2** ($L = \mathbb{Q}(i, \sqrt{6})$). Let $c' = 1$, $c = 6$. We can choose as $u$ the fundamental unit $u = (1+i)\left(-\frac{\sqrt{3}}{\sqrt{2}} - 1\right)$. We have $u\sigma(u) = -1$, $u\tau(u) = -i$. Therefore $a = i\sqrt{6}$, $b = \frac{1+i}{2}$. As in the previous case, $ab\tau(u) = -\sqrt{6}\left(\frac{\sqrt{3}}{\sqrt{2}} - 1\right) < 0$.

**Example 3** ($L = \mathbb{Q}(i, \sqrt{11})$). Let $c' = 1$, $c = 11$. We can choose as $u$ the unit $u = \left(\frac{1+i}{2}\right)(-3-\sqrt{11})$. We have $u\sigma(u) = -1$, $u\tau(u) = -i$. Therefore $a = i\sqrt{11}$, $b = \frac{1+i}{2}$. Again, we have $ab\tau(u) = \sqrt{11}\left(\frac{3-\sqrt{11}}{2}\right) < 0$.

We conclude by giving an example with $c' = 2$ instead of $c' = 1$, with $c = 5$ to get a division algebra.

**Example 4** ($L = \mathbb{Q}(i\sqrt{2}, i\sqrt{5})$). Let $c' = 2$, $c = 5$. The fundamental unit is $u = 3 - \sqrt{10}$. We have $u\sigma(u) = -1$, $u\tau(u) = -1$, $a = i\sqrt{5}$, $b = i\sqrt{2}$, and $ab\tau(u) = -\sqrt{10}(3+\sqrt{10}) < 0$.

**Remark 1.** Ideally, the parameters $u$, $a$ and $b$ ought to be of complex norm 1 in order to have good energy efficiency [1]. Unfortunately, this does not seem to be possible. Indeed, in the case of two imaginary quadratic subfields $K(\sqrt{-c})$ and $K(\sqrt{-c'})$, if $a$ and $b$ have complex norm 1, since the automorphisms $\sigma$ and $\tau$ act as the complex conjugation respectively on $K(\sqrt{-c})$ and $K(\sqrt{-c'})$, we have $a\tau(a) = |a|^2 = 1$ and $b\sigma(b) = |b|^2 = 1$. As a consequence, the condition in Theorem 1 cannot hold and the crossed-product algebra is not a division algebra, since $\forall q \in \mathbb{Q}$, the quaternion algebra $\left(\frac{1,q}{\mathbb{Q}}\right)$ is never a division algebra. Clearly, the case of real quadratic subfields is also hopeless because they do not contain any non-trivial roots of unity.

We finally give an example of a crossed product algebra which is not a division algebra, but provides good shaping:

**Example 5** ($L = \mathbb{Q}(i, \sqrt{3})$). Let $c' = 1$, $c = 3$ as in Example 1, and choose $u = \zeta = \zeta_{12}$. We have $u\sigma(u) = \zeta^8$, $u\tau(u) = -1$. Then $a = 1 + \zeta^8 = \zeta^{10}$ and $b = i$. We have $ab\tau(u) = -1$, so the conditions (13) are satisfied. Since $u$, $a$ and $b$ have complex norm 1, this code provides a very good shaping, although unfortunately it is not full-diversity.

### B. Code optimization

As we have seen, in order to obtain an Alamouti-like structure, the price to pay is to unbalance the energy in the codewords. We now consider the code $C$ in Example 1, and discuss some changes to improve its performance.

*Choice of the basis of $L$:* Recall that an element of $L$ is of the form $x = a_1 + a_2\sqrt{-c'} + a_3\sqrt{-c} + a_4\sqrt{-c'}\sqrt{-c}$, $a_1, a_2, a_3, a_4 \in \mathbb{Q}$. Thus Examples 1-3 and 5 allow QAM encoding, since:

$$x = a_1 + a_2 i + a_3\sqrt{-c} + a_3\sqrt{-c}i = (a_1+a_2 i) + \sqrt{-c}(a_3+ia_4)$$

where both $a_1 + a_2 i$ and $a_3 + ia_4$ are in $\mathbb{Q}(i)$.
In particular, to encode QAM symbols, we can choose any basis of $L$ as a vector space over $\mathbb{Q}(i)$. We will consider two bases, which may not coincide: the basis $\mathcal{B}_1$ of the ring of integers $\mathcal{O}_L$ over $\mathbb{Z}(i)$, and the basis $\mathcal{B}_2 = \{1, \sqrt{-c}\}$. $\mathcal{B}_1$ corresponds to a denser lattice, but we will see that $\mathcal{B}_2$ is more convenient for fast decodability, since it is composed by one purely real and one purely imaginary element.
Let $C_2$ be the version of the code $C$ employing the basis $\mathcal{B}_2 = \{1, \sqrt{3}i\}$. Its drawback is that the basis vectors have unbalanced norms. In order to improve the shaping, we also consider the code $C_3$ with basis $\mathcal{B}_3 = \{2, \sqrt{3}i\}$.

*Renormalization of the parameters $a$ and $b$:* Since $a$ and $b$ are only defined up to a rational constant in (6) and (7), we can choose the normalization in such a way that their complex norm is close to 1, for example we can choose $k = l' = \frac{4}{7}$ in Example 1 (code $C_4$). However, this renormalization affects the minimum determinant of the code; indeed, if $k$ and $l$ are not integers, the code is not contained in the natural order of the crossed-product algebra. A further improvement can be obtained by multiplying the block $C$ and dividing the block $B$ by in (4) by $|a|^{\frac{1}{4}}$; this does not affect the determinant.

## IV. FAST DECODABILITY

In this section we briefly review some facts about decoding complexity. Consider a linear dispersion code encoding $K$ real symbols $a_1, \ldots, a_K$ in a constellation $\mathcal{S}$ such that each codeword is of the form $X = \sum_{i=1}^{K} A_i a_i$ with generator matrices $A_1, \ldots, A_K$. The code is called *conditionally g-group decodable* [8] if there exists a partition of $\{1, \ldots, K\}$ into $g+1$ disjoint subsets $\Gamma_1, \ldots, \Gamma_g, \Gamma^C$ such that

$$b_{lj} = \left\| A_l A_m^H + A_m A_l^H \right\|_F = 0 \quad \forall l \in \Gamma_i, \forall m \in \Gamma_j, i \neq j.$$

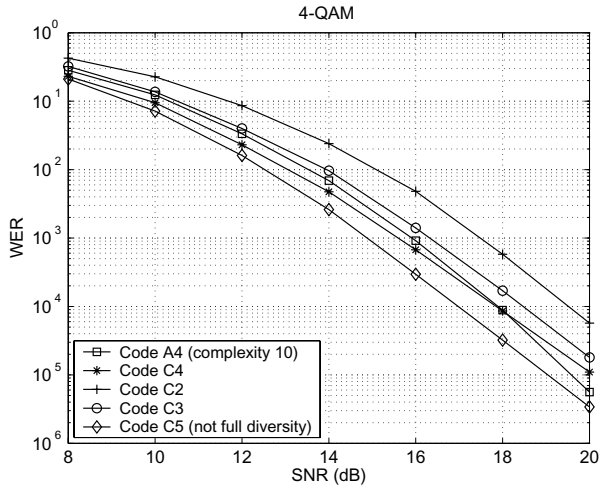

Figure 1. Performance comparison of the proposed codes using 4-QAM constellations with the version of the $A_4$ code with $M^{10}$ decoding complexity.

In this case, the sphere decoding complexity order reduces to $M^{|\Gamma^C|\max_{1\leq i \leq g}|\Gamma_i|}$, where $M$ is the size of $\mathcal{S}$.

For codes of the form (14), using the basis $\mathcal{B}_2$, one can check that the Hurwitz-Radon matrix $B=(b_{ij})$ [3] has the shape

$$\begin{bmatrix} t & t & 0 & 0 & 0 & 0 & 0 & 0 & t & t & t & t & t & t & t & t \\ t & t & 0 & 0 & 0 & 0 & 0 & 0 & t & t & t & t & t & t & t & t \\ 0 & 0 & t & t & 0 & 0 & 0 & 0 & t & t & t & t & t & t & t & t \\ 0 & 0 & t & t & 0 & 0 & 0 & 0 & t & t & t & t & t & t & t & t \\ 0 & 0 & 0 & 0 & t & t & 0 & 0 & t & t & t & t & t & t & t & t \\ 0 & 0 & 0 & 0 & t & t & 0 & 0 & t & t & t & t & t & t & t & t \\ 0 & 0 & 0 & 0 & 0 & 0 & t & t & t & t & t & t & t & t & t & t \\ 0 & 0 & 0 & 0 & 0 & 0 & t & t & t & t & t & t & t & t & t & t \\ t & t & t & t & t & t & t & t & 0 & 0 & 0 & 0 & 0 & 0 & 0 & 0 \\ t & t & t & t & t & t & t & t & 0 & 0 & 0 & 0 & 0 & 0 & 0 & 0 \\ t & t & t & t & t & t & t & t & 0 & 0 & t & t & 0 & 0 & 0 & 0 \\ t & t & t & t & t & t & t & t & 0 & 0 & t & t & 0 & 0 & 0 & 0 \\ t & t & t & t & t & t & t & t & 0 & 0 & 0 & 0 & t & t & 0 & 0 \\ t & t & t & t & t & t & t & t & 0 & 0 & 0 & 0 & t & t & 0 & 0 \\ t & t & t & t & t & t & t & t & 0 & 0 & 0 & 0 & 0 & 0 & t & t \\ t & t & t & t & t & t & t & t & 0 & 0 & 0 & 0 & 0 & 0 & t & t \end{bmatrix}$$

where $t$ denotes any possibly nonzero symbol. Clearly, this code is conditionally 4-group decodable and has complexity order 10. In fact, in order to decode one can list all the possible values for the variables $\{x_9, x_{10}, \ldots, x_{16}\}$ and then minimize separately the Euclidean distance over the pairs of variables $\{x_1, x_2\}, \{x_3, x_4\}, \{x_5, x_6\}$ and $\{x_7, x_8\}$.

When using the basis $\mathcal{B}_1$, one can see that the complexity order is 12 (details are omitted for lack of space).

## V. SIMULATIONS

Figure 1 shows the performance of the proposed codes using 4-QAM, compared to the $O(M^{10})$ decoding complexity version of the "$A_4$ code" in [12] at the same spectral efficiency. The code $C_2$ loses about $1.6\,\mathrm{dB}$ with respect to the $A_4$ code at the WER of $10^{-4}$. Thanks to its more balanced basis, the code $C_3$ improves the performance by $1\,\mathrm{dB}$. Finally, the renormalized version $C_4$ has better performance than the $A_4$ code in the low SNR regime (up to SNR=18) but does not work so well at high SNR due to its small minimum determinant.

Due to its excellent shaping, the code $C_5$ from Example 5 performs surprisingly well at low SNR, and in fact it outperforms the $A_4$ code by $0.9\,\mathrm{dB}$ at the WER of $10^{-4}$, even though the error rate will eventually be worse in the high SNR regime since the code is not full rank.

## VI. CONCLUSIONS

We proposed a new family of full-rate, non-vanishing determinant $4 \times 4$ MIDO codes based on cross-product algebras over $\mathbb{Q}$ with ML decoding complexity order $O(M^{10})$. Simulation results show a performance from similar up to a 1 dB gain compared to the best previously known code with the same complexity order [12]. While the latter requires real PAM signal constellations, our codes further have the advantage of being suitable for QAM complex modulation.

## ACKNOWLEDGMENTS

This work was partly done while L. Luzzi was visiting the Division of Mathematical Sciences, Nanyang Technological University, Singapore.
The research of F. Oggier is supported in part by the Singapore National Research Foundation under Research Grant NRF-RF2009-07 and NRF-CRP2-2007-03, and in part by the Nanyang Technological University under Research Grant M58110049 and M58110070. The research of L. Luzzi is supported by the Supélec Foundation and by NEWCOM++.